\documentstyle[epsf,twocolumn,prb,aps,amssymb]{revtex}
\begin{document}
{

\draft
\title{Can Electric Field Induced Energy Gaps In Metallic Carbon Nanotubes? }
\author{ Xin Zhou$^{1}$, Hu Chen$^{2}$ and Ou-Yang 
Zhong-can$^{1,2}$}
\address{$^{1}$Institute of Theoretical Physics, The Chinese 
Academy of Sciences, P. O. Box 2735, Beijing 100080, China
\\ $^{2}$Center for Advanced Study, Tsinghua University, Beijing 100084,
 China}
\date{\today}

\maketitle

\begin{abstract}
The low-energy electronic structure of metallic single-walled carbon 
nanotube (SWNT) in an external electric field perpendicular 
to the tube axis is investigated. Based on tight-binding approximation, 
 a field-induced energy gap is found in all (n, n) SWNTs, and the 
gap shows strong dependence on the electric field and the 
size of the tubes.
We numerically find a universal scaling that the gap is a function of 
the electric field and the radius of SWNTs, and the results
are testified by the second-order perturbation theory in weak field limit. 
Our calculation shows the field required to induce a $0.1$ ${\rm eV}$ 
gap in metallic SWNTs can be easily reached under the current experimental 
conditions. It indicates a kind of possibility to apply nanotubes to 
electric signal-controlled nanoscale switching devices.
\end{abstract}

}

The prospect of nanoscale electronic devices has engaged great 
interest, because it could lead to conceptually new
miniaturization strategies in the electronics and computer 
industry. Single-walled carbon nanotubes (SWNTs) can be used
as nanoscale devices~\cite{Ptody} 
due to their extraordinarily small diameter and versatile 
electronic properties~\cite{Hamada}. It is suggested that individual SWNT may act as
 devices such as field-effect transistors (FET)~\cite{Tans1}, 
single-electron-tunneling transistors~\cite{Tans2,Bockrath}, 
rectifiers~\cite{Yao,Fuhrer}, or p-n junctions~\cite{Leonard}. The
most exciting expectancy lies in the devices fabricated on a single tube~\cite{[9]}.

In recent years, the interplay between mechanical deformation and 
electrical properties of SWNTs have been extensively 
studied~\cite{[9],Bezryadin,Crespi,Kane}. Among them, some structural
deformations such as twisting, bending, stretching, and topological
defects are not compatible with desirable stable contacts for reversibly controllable devices~\cite{Lammert}. Very recently, Tombler {\it et al.}~\cite{Tombler,Liu} used an atomic force
microscope tip to manipulate a metallic SWNT, leading to a reversible 
two-order magnitude change of conductance, and Lammert 
{\it et al.}~\cite{Lammert} applied a uniaxial stress to squash 
SWNTs and detect a similar reversible metal-insulator (M-I) transition. Since the tube ends do not need to move and they are easily controlled, the studies
pointed to the possibility that a metallic SWNT may be used as an 
ultrasmall electromechanical switch. 

It is also well known that a magnetic field can change the conductance of 
carbon nanotubes~\cite{Ajiki,Lu,Roche}. The magnetic 
field either parallel or perpendicular to the axis can change
the low-energy electronic properties of the tubes. 
However, a possible electric field-controlled M-I transition 
 are more exciting because of its easy implementation in the actual 
 applications. Now, a question remains: 
Can electric field change the electronic properties of a tube?
  
In previous studies on electronic transports~\cite{Datta}, 
since the bias voltage is a slow-change variable in the range of the 
primitive unit cell, the possible change of the energy-band structure
induced by the bias voltage is neglected.
The controlled potential, such as the gate voltage in the case of 
FET~\cite{Tans1}, does not induce a voltage drop 
in the direction perpendicular to the tube axis, and it only shifts 
Fermi level or changes the carrier concentration. 
In the literature, according to our knowledge, 
it has not been studied on that the properties of the longitudinal 
electronic transports are changed by a transverse electric field. 
In fact, because of the small diameter of SWNTs ( about $1$ ${\rm nm}$ ), 
it is easy to exert a strong electric field 
 ( $|{\vec E}| \sim {\rm V/ nm} \sim 10^{8 - 9}$ ${\rm V/m}$ )
 perpendicular to the axis. In $(n,n)$ metallic SWNT, the electrons
near Fermi energy are nonlocal in the circumference of the tube since their
circumference Fermi wavevector is zero~\cite{Hamada}, the classic wave-package 
approximation in slow-change voltage may be not suitable in the presence of the 
strong transverse electric field. 
 The ${\rm V/ nm}$ order electric field is enough to 
 obviously break the rotational symmetry about the tube 
 axis, and create new interband and inter-wavevector coupling, 
 which may change the low energy electronic 
 properties of SWNTs, and hence affect the electronic transport. 
In the other hand, the field is still $2 \sim 3$ orders less than the
atomic interior electric field, can be treated as perturbation.

\begin{figure}
\centerline{\epsfxsize=10cm \epsfbox{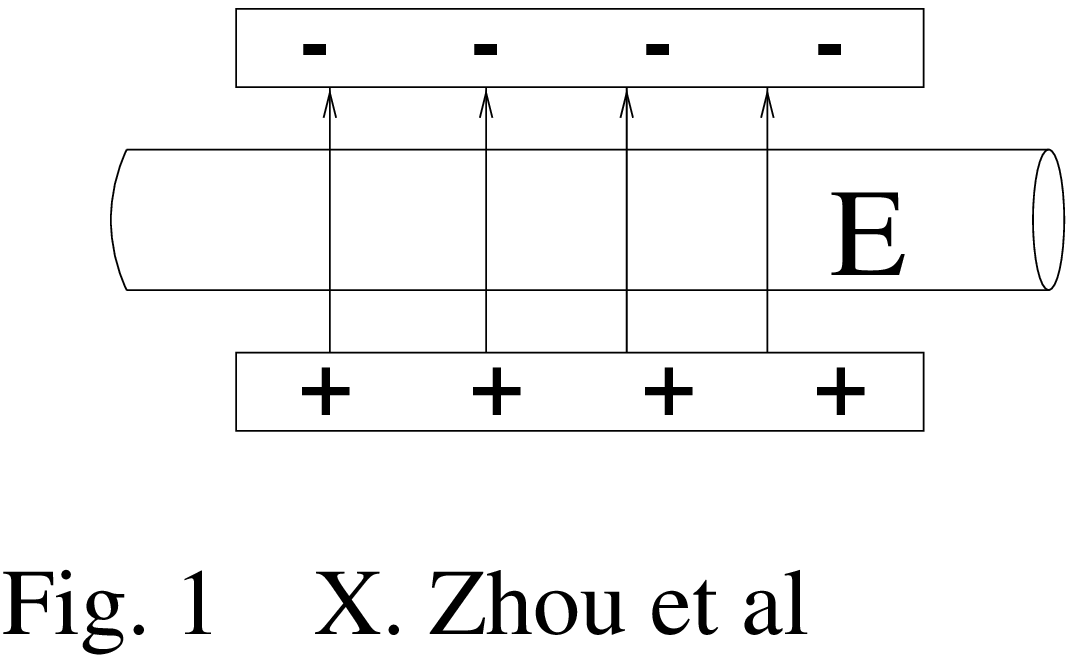}}
\caption{A uniform external electric 
field perpendicular to the axis of SWNT. 
\label{fig1}}
\end{figure}

 In this Letter, based on a tight-binding (TB) model, we 
 calculate the low-energy electronic structure of SWNTs in an external 
 electric field perpendicular to the tube 
  axis (see Fig.\ref{fig1}). The result shows obviously valuable effects: 
 (1) The electric field can always induce an energy gap in $(n, n)$ metallic SWNTs; 
 (2) There is a maximal gap strongly dependent on the 
 radius of the tubes;
 (3) A universal scaling is found for the gap as 
 a function of the field and the size of the tubes;
 (4) Using the dielectric function of tubes obtained by Benedict 
 {\it et al.}~\cite{Benedict}, we find
  the magnitude of the required electric field falls into the range 
  of current experimental conditions, therefore the possibility 
  of applying SWNTs to the 
 electric-field-controlled nanoscale switching devices.

 The nearest-neighbor TB approximation has been 
 used successfully for calculating the electronic structure of 
 graphite sheet and nanotubes~\cite{Hamada}, and the polarization of 
 SWNTs~\cite{Benedict}. 
 For the low-energy electronic properties of tubes in the
 presence of electric field, we will 
 only use the $\pi$-electron single-orbital TB approximation~\cite{zhouxin1}.
 
 In the presence of a transverse electric field ${\mathbf E}$, 
 there is an additional coupling between the nearest-neighbor atoms, 
 reflected in the TB Hamiltonian
 \begin{equation}
  {\cal H}={\cal H}_0 + {\cal H}_1 ,
 \end{equation}
 where ${\cal H}_0$ is the unperturbed Hamiltonian, and 
 ${\cal H}_1 = e V({\mathbf r})$. Here $e$ is the charge of an electron, 
 and $V({\mathbf r})$ is the electrostatic potential of the total electric 
 field which equals to the sum of an external field ${\mathbf E}_{0}$ and the
 polarized field induced by ${\mathbf E}_{0}$. For a uniform 
 field~\cite{zhouxin2}, the potential in the cylindrical surface of SWNT is 
 \begin{equation}
 V({\mathbf r})=-{V_0} \cos \phi, 
 \end{equation} 
 where $V_0= R E $ is the transverse voltage drop of 
 the tube, and $\phi$ is the azimuth of the 
 cylinder. Here $R$ is the radius of SWNT, and $ E $ is 
 the total electric field strength.
 
 The main role of the electrostatic potential 
 is to change the electronic energy of the $i$th carbon atom
 by $e V({\mathbf R}_{i})$, where ${\mathbf R}_{i}$ is the position vector 
 of the $i$th atom. The hopping correction due to the electric field between 
 site $i$ and site $j$ 
 is very small, about the order of $s$. Here $s$ is
 the overlap integral of two nearest neighbor atom $i$ and 
 atom $j$,
  $s=\int{{\phi}_{i}^{*}({\mathbf r}) {\phi}_{j}({\mathbf r})
  d^3 {\mathbf r}} \approx 0.129$~\cite{Zhou1}, where 
  ${\phi}_{i}({\mathbf r})$ is the $\pi$-electron wavefunction of the $i$th
  atom. 
 The perturbed Hamiltonian matrix elements are:
 \begin{eqnarray}
 &&<{\phi}_{i}|{\cal H}_{1}
 |{\phi}_{i}> = e V({\mathbf R}_{i}), \label{AA}\\
 &&<{\phi}_{i}|{\cal H}_{1} |{\phi}_{j}> \approx 
       e s V({\mathbf R}^{c}_{ij}),
      \label{AB}
 \end{eqnarray} 
 where ${\mathbf R}^{c}_{ij}= ({\mathbf R}_{i} + {\mathbf R}_{j})/2$ is the 
 position of the mass center of the two nearest neighbor $i$ atom and $j$ atom.
 Eq.(\ref{AA}) is precise because of the symmetry of wavefunction, 
 but in Eq.(\ref{AB}), we have neglected the electrostatic potential 
 change in the overlapping range of electronic clouds. 
 After gaining these Hamiltonian matrix elements, we can easily calculate the
 electronic structure of nanotube in the presence of electric field.
 
\begin{figure}
\centerline{\epsfxsize=10cm \epsfbox{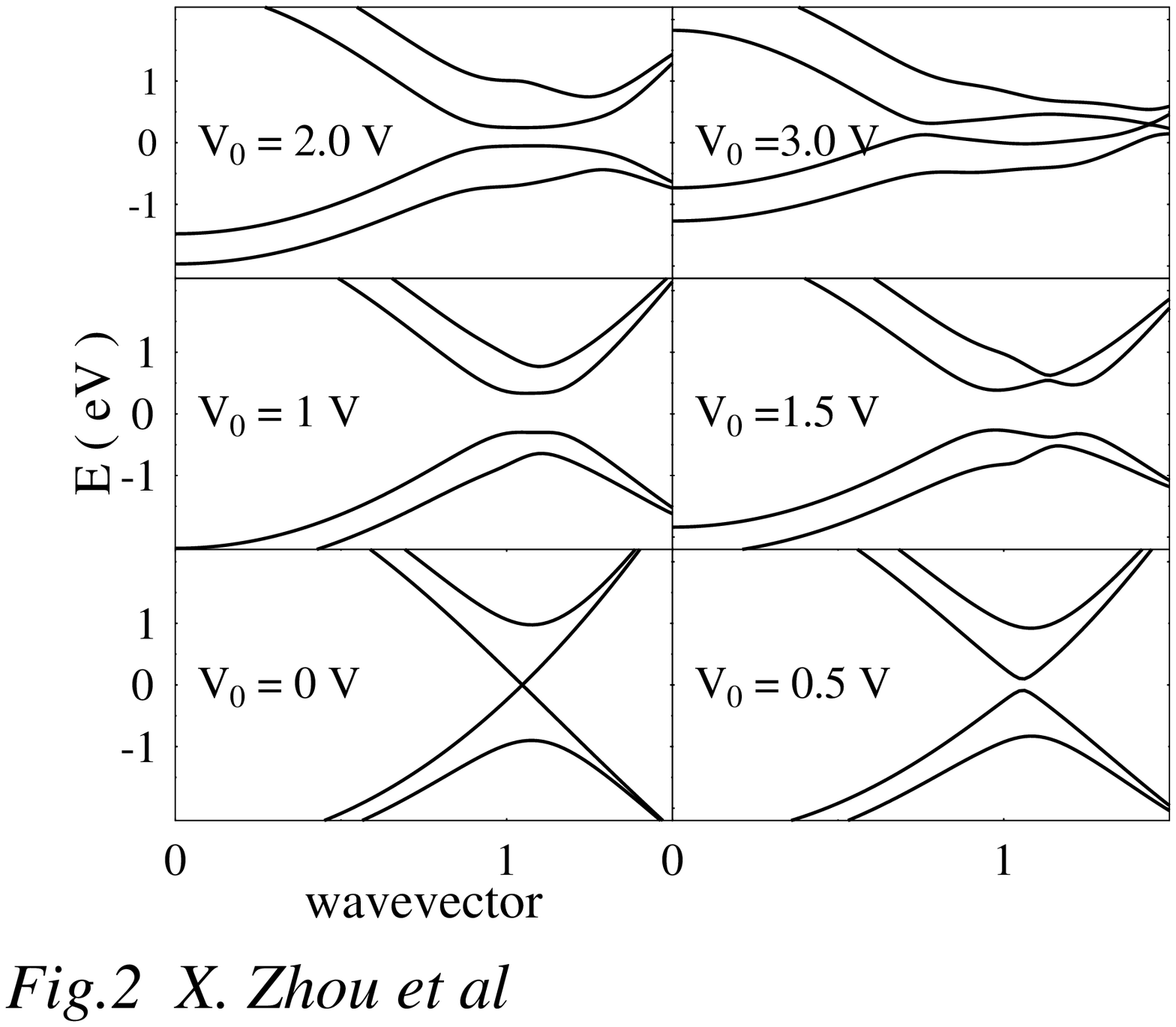}}
\caption{The energy bands of $(10, 10)$ SWNT in the vicinity of   
Fermi level under the application of a transverse electric 
field $V_{0}$. (10, 10) tube is metal at $V_{0}= 0$ ${\rm V}$, and it
is a semiconductor in the weak field range. 
\label{fig2}}
\end{figure}

 Fig.\ref{fig2} shows our result on the electronic energy bands of 
 (10, 10) tube in
 the presence of electric fields with different magnitudes. When
 $V_{0} = 0.5$ ${\rm V}$, a gap of $E_g \sim 0.3$ ${\rm eV}$ is 
 found at $K_0$, the Fermi wavevector in zero field. As $V_{0}$
 increases, the gap increases, and when $V_{0}=1.5$ 
 ${\rm V}$, we find that the bands structure is obviously deformed. 
 It is surprising to find that the gap decreases as  
 $V_{0}$ increases further. When $V_{0} =3.0$ ${\rm V}$, 
 the zero gap is found, but the Fermi point dramatically moves to a 
 different position from $K_0$. The results reveal that a controlled 
 electric field of $1$ ${\rm V/nm}$ order can obviously affect the 
 transport properties of $(10,10)$ tube. 
 
\begin{figure}
\centerline{\epsfxsize=10cm \epsfbox{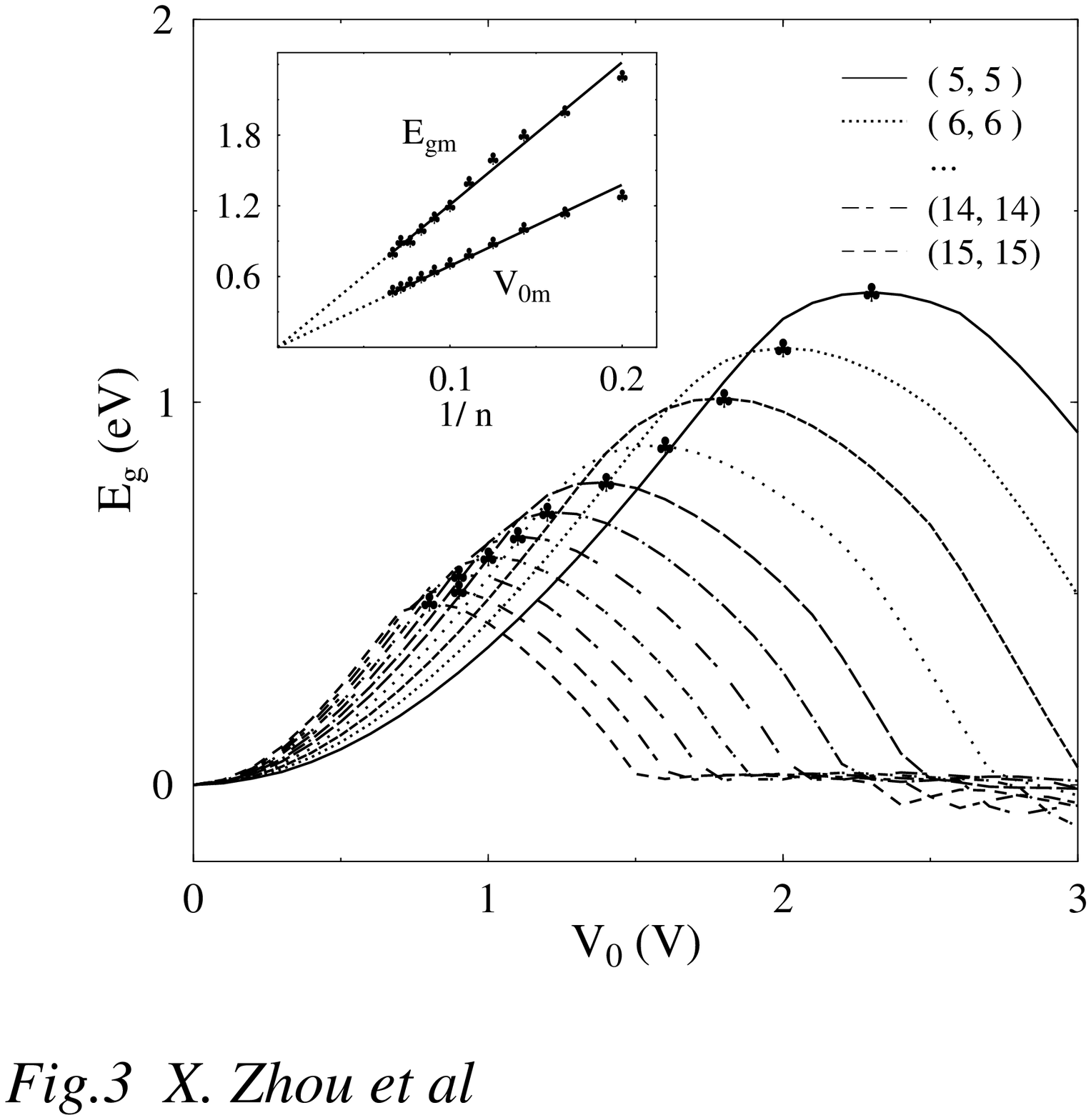}}
\caption{Field-induced gap of $(n, n)$ 
tubes versus the field. From the top to the bottom, 
the tube parameter $n$ increases from $5$ to $15$. The clubs 
denote the position of the maximum gap point. In the inset, both the 
maximum gap $E_{gm}$ and $V_{0m}$ ( see Text )
are found to be proportional to $1/n$
The lines are fitting results.
\label{fig3}}
\end{figure}

\begin{figure}
\centerline{\epsfxsize=10cm \epsfbox{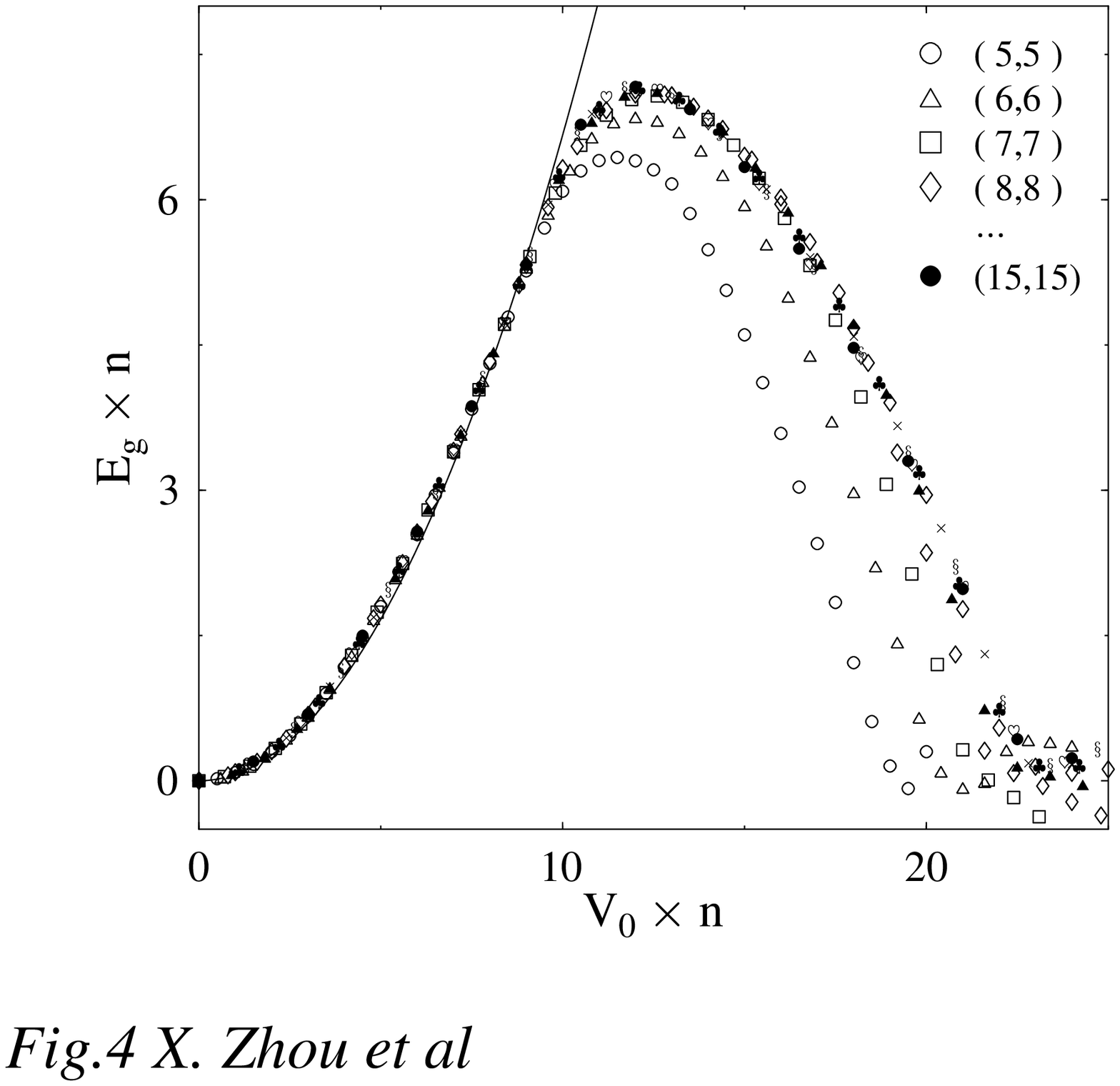}}
\caption{Universal scaling are found for the gap as a
function of $V_{0}$ in different $(n, n)$ tubes. At 
weak field, the data of all tubes are very much consistent with the scaling 
relation $ n E_g = {\lambda} (n e V_{0})^2 $, as expected by the 
second-order perturbation theory. The line is the fitting result. 
Except for (5, 5) and (6,6) 
tubes, the universal scaling is satisfied well up to strong field region. 
\label{fig4}}
\end{figure}

 To probe the above effect in general, we performed the computation 
 for a series of $(n,n)$ tubes. Fig.\ref{fig3} shows  
 the gap as a function of the electric field in $(n, n)$ 
 tubes, where $n$ is from $5$ to $15$. From the figure, we find the 
 determined effect: The electric field can always induce a gap, i.e. a 
 metal-insulator transition, in $(n, n)$ tubes, 
 and the size of the gap strongly depends on the magnitude of 
 the transverse field and
 the tube parameter $n$. For any $(n, n)$ tubes, the gap first 
 increases with increasing field, and reaches a maximal
 value $E_{gm}$ at the $V_{0}=V_{0m}$, then drops again. Both the 
 maximal gap $E_{gm}$ and the corresponding  
 $V_{0m}$ are approximately proportional to 
 $1/n$, and hence inverse proportional to the radius of tubes, i.e.,
 \begin{eqnarray}
  E_{gm} \approx 6.89\ {\rm eV}/n, \label{Egm}\\
  V_{0m} \approx 12.09\ {\rm V}/n. \label{em}
 \end{eqnarray}
 The finding, which is shown in (\ref{Egm}) and (\ref{em}) that the maximal gap 
 $E_{gm}$ fastly decrease as the 
 size of conductor increases, indicates the electronic structure change is very
 small in the large materials even though at very strong electric field. It 
 might be the reason why people have not yet recognized the effect in previous 
 studies on large conductors. In the strong
 field range, Fig.\ref{fig3} shows a more complicated field dependence of
 the gap which can even be negative.
 The field dependence of the gap is quite similar for tubes with
 various radii, which invokes us to scale both $E_g$ 
 and $V_{0}$ up $n$ times their original values.
 The obtained results are shown in Fig.\ref{fig4}. From it we do find 
 the scaled gap to be a universal function of the scaled electric 
 field for all $(n,n)$ tubes. In the weak field range,
 for all calculated eleven $(n, n)$ tubes there exists a simple relation:
  $n E_g = \lambda (n e V_{0})^2$,
 where ${\lambda}$ is a constant, about $0.07$ ${\rm (eV)^{-1}}$.
 In the middle field range, except for a few small-radius tubes 
 such as $(5, 5)$ and $(6, 6)$ tubes, the universal scaling law still holds. 
 
 To understand the above scaling relation, we use perturbation
 theory to calculate the field-induced gap in weak field limit. 
 The first-order perturbation approximation only causes shift in the 
 Fermi level, showing no contribution to the gap. Calculating up 
 to the second-order perturbation 
 at $K_0$ point, we obtained the following analytic result
 \begin{equation}
  E_g \approx \frac{\sqrt{3}}{2 \pi \gamma} n e^2 {V_{0}}^2,
  \label{perturb}
 \end{equation}
 where $\gamma$ (=$3.033$ eV) is the hopping parameter in the absence of 
 the electric field~\cite{Zhou1,White}. The contribution of the 
 overlap integral $s$, which is very small, is neglected. Obviously, 
 the second-order perturbation calculation gives almost the same 
 scaling relation as the numerical results in the weak field, though 
 the obtained $\lambda \approx 0.09$ ({\rm eV})$^{-1}$ is slightly 
 larger than the
numerical result $0.07$ ({\rm eV})$^{-1}$. In the 
 strong field range, since the Fermi wavevector is away from $K_0$, 
 the perturbation theory becomes not suitable.
 In fact, the weak field range may be more compatible with the 
 practical application. In order to open a $0.1$ ${\rm eV}$ gap in 
 the energy bands of $(n, n)$ 
 tubes, $n$ must be smaller than 68 according to eq.(\ref{Egm}), and 
 the required electric field is,
 \begin{equation}
 E = \frac{2 \pi}{3 r_0 e} \sqrt{\frac{E_g}{\lambda}} n^{-\frac{3}{2}},
 \end{equation}
 where $r_0$ (=$1.42$ ${\rm \AA}$) is the bond length of carbon atoms in 
 SWNT. Therefore, for example for a $(10,10)$ tube, the required field is about 
 $5 \times 10^8$ ${\rm V/m}$, and 
 for a $(60,60)$ tube, it is about $3 \times 10^7$ ${\rm V/m}$.
 Since the total field $\mathbf E$ equals
 to the sum of the external field ${\mathbf E}_0$ and the  
 of the polarized field, we have ${\mathbf E}= {\mathbf E}_0 / {\epsilon}$, where 
 $\epsilon$ is the dielectric function. Under the homogeneous polarization 
 approximation, Benedict {\it et al.}~\cite{Benedict} calculated the 
 $\epsilon$ of some tubes within random phase approximation. Based on their 
 results, we have~\cite{zhouxin3},
 \begin{equation}
 {\epsilon} = 1 + 5.2 \frac{R^2}{(R + \delta R)^{2}},
 \end{equation}
 where $R$ is the radius of tubes. Benedict gave $\delta R$ about
 1.2 $\AA$.
 So the required external field is about 6 times of the calculated total 
 field $\mathbf E$. 
 Since such a magnitude of the electric field can be easily reached by the 
 current experimental conditions, we wish the above prediction 
 can be checked in near future.
  


In summary, we have proposed an electric-field-induced M-I
transition in $(n, n)$ SWNTs for the first time. The universal 
relationship between the gap and the electric field has been 
obtained in SWNTs by using the TB model. The results support the argument that SWNTs can be applied as nanoscale electric signal-controlled switching devices.

 The authors would like to thank Prof. H.-W. Peng, Prof. Z.-B. Su and 
 Dr. H.-J. Zhou 
 for many discussions on the results. The numerical calculations
 were performed partly at ITP-Net and partly at the State Key Lab. 
 of Scientific and Engineering Computing.

\begin{thebibliography}{99}
\bibitem{zhouxin1} The details of the TB approximation in the precense of 
electric field can be found in Refs.~\cite{Benedict}, the coupling 
between $\sigma$- and $\pi$- electrons induced by the curvature effects of 
tubes is very small.
\bibitem{zhouxin2} We suposed the polarization of tubes is 
uniform~\cite{Benedict}, the a uniform external field induce a uniform polarized
field.
\end{thebibliography}
\end{document}